\documentclass[useAMS,usenatbib]{mn2e}

\usepackage{graphicx}

\title[Simulating a faint gamma-ray burst population]{Simulating a faint gamma-ray burst population}
\author[D. M. Coward]{D. M. Coward\thanks{E-mail:
coward@physics.uwa.edu.au} \\School of Physics, University of
Western Australia, M013, Crawly WA 6009, Australia}
\begin{document}

\date{\today}

\pagerange{\pageref{firstpage}--\pageref{lastpage}} \pubyear{2004}

\maketitle

\label{firstpage}

\begin{abstract}
There have now been three supernova-associated $\gamma$-ray bursts
(GRBs) at redshift $z<0.17$, namely 980425, 030329, and 031203,
but the nearby and under-luminous GRBs 980425 and 031203 are
distinctly different from the `classical' or standard GRBs. It has
been suggested that they could be classical GRBs observed away
from their jet axes, or they might belong to a population of
under-energetic GRBs. Recent radio observations of the afterglow
of GRB 980425 suggest that different engines may be responsible
for the observed diversity of cosmic explosions. Given this
assumption, a crude constraint on a luminosity function for faint
GRBs with a mean luminosity similar to that of GRB 980425 and an
upper limit on the rate density of 980425-type events, we simulate
the redshift distribution of under-luminous GRBs assuming BATSE
and \textit{Swift} sensitivities. A local rate density of about
0.6\% of the local supernova Type Ib/c rate yields simulated
probabilities for under-luminous events to occur at rates
comparable to the BATSE GRB low-redshift distribution. In this
scenario the probability of BATSE/HETE detecting at least one GRB
at $z<0.05$ is 0.78 over 4.5 years, a result that is comparable
with observation. {\it Swift} has the potential to detect 1--5
under-luminous GRBs during one year of observation.
\end{abstract}


\section{Introduction}

Discovery of GRB 980425, associated with the very nearby supernova
(SN) 1998bw (at $z=0.0085$, corresponding to about 40 Mpc),
heralded a new era in understanding the origin of GRBs
\citep{Galama1,Galama2}. Recently, the detection of spectroscopic
features in the light curve of GRB 030329, similar to those seen
in SN 1998bw \citep{stanek03,Hjorth03}, has strengthened the SN
1998bw/GRB 980425 association. These observations support the
`collapsar' model \citep{zang03} in which a Wolf-Rayet progenitor,
possibly in a binary system, undergoes core collapse, producing a
compact object surrounded by an accretion disk, which injects
energy into the system and thus acts as a `central engine'. The
energy extracted from the system gives rise to a Type Ib/c SN
explosion and drives collimated jets along the progenitor rotation
axis, producing a prompt GRB and afterglow emission--see the
review by Zhang \& M\'esz\'aros (2004).

Discovery of the GRB-SN association was an important breakthrough,
but GRB 980425 had an unusually low luminosity---it was
under-luminous in $\gamma$-rays by three orders of magnitude
compared to `classical' GRBs. It was suggested that it could be a
very rare event and not a member of the classical GRB population.
This explanation seems unlikely given the discovery of the
under-luminous and nearby GRB 031203 ($z=0.105$), the first
analogue of GRB 980425.

There are presently three GRBs (980425, 030329, and 031203)
definitely associated with extremely energetic Type Ib/c SNe
\citep{Proch04,mal04}, all occurring at $z<0.17$. GRB 030329 is
classified as classical in the context of energy emission, but was
relatively close at $z=0.17$. However, it is difficult to
reconcile the under-luminous GRBs 980425 and 031203 with the
classical population. One simple explanation is that
under-luminous bursts are GRBs observed away from the jet axis.

\cite{g04} argue that a unified picture can only be obtained by
using a luminosity function (LF) that includes all luminosities
down to that of GRB 980425, so that the probability of observing
the three low-$z$ events is non-negligible.  They show that for
GRBs 980425, 030329, and 031203 to belong to the classical burst
population, the LF must be a broken power-law. This is an
attractive proposal in that GRBs 980425 and 031203 can be
explained within the bounds of currently popular GRB progenitor
models by extending the LF to accommodate GRB 980425. They
calculate that if this is the case, no bright burst within
$z=0.17$ should be observed by a HETE--like instrument within the
next $\sim 20$ yr.

\section{Anomalous GRBs}
\subsection{Evidence for intrinsically sub-energetic events}
\cite{sod04} argue that if GRB 031203 was observed `off axis',
then the radio afterglow should brighten as the ejecta slows down,
but they did not observe any re-brightening. They find that the
afterglow is faint, indicating that the explosion was
under-energetic. Similarly, there is no evidence of re-brightening
for GRB 980425, despite radio calorimetry since 1998. \cite{sod04}
conclude that GRBs 980425 and 031203 were intrinsically
sub-energetic events.

If SN 1998bw was a rare and unusually sub-energetic SN distinct
from local SNe and GRBs, \cite{sod04} claim the characteristics of
SN 1998bw/GRB 980425 are not a result of the observer's viewing
angle but of the properties of its central engine. SN 1998bw was
an engine-driven explosion \citep{li99}, in which 99.5\% of the
kinetic energy ($\sim 10^{50}$ ergs) was coupled to relativistic
ejecta of Lorentz factor 2 \citep{kulk98}, while a mere 0.5\% went
into the ultra-relativistic flow. In contrast, `classical' GRBs
couple most of their energy into $\gamma$-rays. \cite{berg03b}
claim that the observed diversity of cosmic explosions (SNe, x-ray
flashes (XRF), and GRBs) could be explained with a standard energy
source but with a varying fraction of that energy injected into
relativistic ejecta. Different engines may be responsible for the
observed diversity of cosmic explosions, implying that classical
GRBs represent one class of event, one in which $\gamma$-rays
channel most of the energy away from a central engine.

It is evident that SN 1998bw could be a member of a distinct class
of SN explosions. But how rare is SN 1998bw in the context of Type
Ib/c SNe and classical GRBs? \cite{berg03a} carried out a
systematic program of radio observations of Type Ib/c SNe using
the Very Large Array to place the first constraint on the rate
density of SN 1998bw type events. Of the 33 SNe observed from late
1999 to the end of 2002, they conclude that the fraction of events
similar to SN 1998bw is at most 3\%. Furthermore they find, by
comparison of the SN radio emission to that of GRB afterglows,
that none of the observed SNe could have resulted from a classical
GRB.

\subsection{Evidence for an off-axis model for GRB 031203 }
The evidence that GRB 031203 was an intrinsically faint and nearly
spherical explosion is not widely accepted. \cite{Rameriz04}
disagree with this interpretation and argue, from models fitted to
the observed x-ray light curve and the radio afterglow, that GRB
031203 was a classical GRB viewed off axis. They find that most
spherical models under-predict the x-ray flux at late times by at
least two orders of magnitude, and prefer to interpret GRB 031203
as a highly collimated GRB viewed off-axis.

For the case of an observer located outside the jet aperture,
$\theta_\mathrm{obs}>\theta_0$, the prompt GRB emission and its
early afterglow are considerably weaker than for on axis,
$\theta_\mathrm{obs}<\theta_0$. An observer at
$\theta_\mathrm{obs}>\theta_0$ sees a rising afterglow light curve
at early times, which approaches that seen by an on-axis observer
at late times. The emission remains low until the cone of the beam
intersects the observer's line of sight.

In the off-axis jet scenario, with viewing angle $\theta_\mathrm
{obs}\sim 2\theta_0$, GRB 031203 can be modelled as a GRB viewed a
few degrees outside of a conical jet. \cite{Rameriz04} show that
if GRB 031203 is modelled as an off-axis observation, its energy
emission in $\gamma$-rays is about 10$^{53}$ ergs, consistent with
classical GRBs.

The off-axis model of \cite{Rameriz04} used to explain the weak
afterglow of GRB 031203 casts some doubt on the claim that it is
an intrinsically weak event. Furthermore, because GRBs 031203 and
980425 share a common deficit in $\gamma$-ray emission, it is
possible that these `outliers' are a result of off-axis
observations. The possibility that GRBs may represent a class of
cosmic explosions with a broad range of energies provides the
basis for the following simulation.

\subsection{A population of under-luminous GRBs}
Given the potential of the \textit{Swift} satellite, a
multi-wavelength GRB observatory (http://swift.gsfc.nasa.gov/)
launched on 2004 November 20, to localize hundreds of GRBs,
observations of events similar to GRB 980425 will provide new
insight into GRB progenitor populations. This is strong motivation
to constrain the probability of detecting similar GRB-SNe assuming
a simple, if highly uncertain, model. The first upper limits on
the rate density of Type Ic SNe associated with GRBs are being
made, based on the presence of jetted emissions (e.g. Berger et
al. 2003a). Given such upper limits, one can at least provide
further constraints on GRB progenitor populations using continued
satellite observations. Furthermore, we provide an example to
demonstrate how an under-luminous GRB population would manifest
during the BATSE observation period and the era of \textit{Swift}.

Firstly, we assume GRB 980425 is a `typical' member of a class of
relatively rare GRB (compared to classical GRBs). Based on the
luminosity of GRB 980425, the mean luminosity of this population
is about 3 orders of magnitude less than that of the classical
population. Secondly, the local rate density must be less than 3\%
of the Type Ib/c SN rate. Another observational constraint is
based on the observed rate of GRB 980425 type events out to
$z=0.0085$ assuming a 4.5 yr (BATSE) observation period; the
probability of occurrence of this very nearby GRB must be
compatible with the BATSE GRB distribution. Finally, we assume the
$\gamma$-ray emissions are isotropic (not beamed) based on the
radio observations of the afterglows of GRBs 980425 and 031203
\citep{sod04}. With these constraints, we simulate the observed
GRB distribution for BATSE and \textit{Swift} sensitivities.

\section{The GRB luminosity function}
The GRB LF, together with the flux sensitivity threshold of an
instrument, determine the fraction of all GRBs potentially
detectable with that instrument:
\begin{equation}\label{grbrate}
\psi_{\rm GRB}(z)= S_d\int_{L_{\rm lim}(z)}^{\infty}p(L) {\rm d}L
\;,
\end{equation}
where $\psi_{\rm GRB}(z)$ is the GRB rate scaling function, $S_d$
is the fraction of sky that the detector scans, and $p(L)$ is the
GRB LF with $L$ the intrinsic luminosity in units of photons
s$^{-1}$. With $f_{\rm lim}$ denoting the instrumental flux
sensitivity threshold, in photons s$^{-1}$ m$^{-2}$, the minimum
detectable luminosity can be expressed as a function of redshift
by $L_{\rm lim}(z)= 4 \pi {D_{\mathrm L}}^{2}(z) f_{\rm lim}$,
with $D_{\mathrm L}(z)$ the luminosity distance.

Most models for the classical GRB LF are based on the
luminosity--redshift relation \citep{schaef01}. But the models are
biased by the redshift sample and the sensitivity limit related to
the redshift estimate. This limit has to be lowered at least by an
order of magnitude to encompass the complete range of
luminosities. \cite{firmani04} show that by jointly fitting to the
observed differential peak-flux and redshift distributions, the
best fit for the LF takes a form that evolves weakly with
redshift. However, there is no consensus on the form of LF for
classical GRBs; possibilities include a single power law, a double
power law and a log-normal distribution.

For an under-luminous population of GRBs modelled on the single
GRB 980425, the choice of LF is so uncertain that the form of the
function is somewhat arbitrary. Nonetheless, for definiteness and
for comparison, we use log-normal distributions for the classical
GRBs and an under-luminous population, both with observation-based
statistical moments that fit the data:
\begin{equation}\label{PL}
p(L)= \frac{{\rm e}^{-\sigma^{2}/2}}{\sqrt{2\pi \sigma^{2}}} \exp
\bigg\{-\frac{[\ln(L/L_{0})]^{2}}{2\sigma^{2}}\bigg\}\frac{1}{L_{0}}
\;,
\end{equation}
where $\sigma$ and $L_{0}$ are the width and average luminosity,
respectively. We take $\sigma=2$ and $L_{0}=2\times 10^{56}$
s$^{-1}$ for the classical GRBs, with $f_{\rm lim}=0.2$ and 0.04
photons s$^{-1}$ cm$^{-2}$ for BATSE (classical GRBs) and {\it
Swift} respectively, and take about 0.1 for $S_d$ \citep{g04}.
Assuming that GRB 980425 is representative of an under-luminous
population, we take $L_{0}=2\times 10^{53}$ s$^{-1}$, three orders
of magnitude less than for the classical GRBs.

\section{GRB rates and  detection probability}
One can express the differential GRB rate in the redshift shell
$z$ to $z+{\mathrm d}z$ as
\begin{equation}\label{drdz}
\mathrm{d}R =  \psi(z) \frac{\mathrm{d}V}{\mathrm{d}z}\frac{r_0
e(z)}{1+z} \mathrm{d}z \,,
\end{equation}
\noindent where $\mathrm{d}V$ is the cosmology-dependent co-moving
volume element and $R(z)$ is the GRB event rate, as observed in
our local frame, for sources out to redshift $z$. Source rate
density evolution is accounted for by the dimensionless evolution
factor $e(z)$, which is normalized to unity in the present-epoch
universe $(z=0)$, and $r_0$ is the $z=0$ rate density. The $(1 +
z)$ factor accounts for the time dilation of the observed rate by
cosmic expansion.

We assume a `flat-$\Lambda$' cosmology with $\Omega_{\mathrm
m}=0.3$ and $\Omega_{\mathrm \Lambda}=0.7$ for the present-epoch
density parameters, and take \mbox{$H_{0}=70$ km s$^{-1}$
Mpc$^{-1}$} for the Hubble parameter at $z=0$. The star formation
rate (SFR) model SF2 of \cite{mad01} is re-scaled to this
cosmology and converted to a normalized evolution factor $e(z)$.
This SFR model levels off to an essentially constant rate density,
of order 10 times the $z=0$ value, at $z>2$; a star formation
cutoff at $z=10$ is assumed.

For the classical GRBs we use $r_0 = 0.9$ yr$^{-1}$ Gpc$^{-3}$,
obtained by scaling the all-sky Universal rate to 692 yr$^{-1}$,
the value implied by the
GUSBAD\footnote[1]{http://www.astro.caltech.edu/$\sim$mxs/grb/GUSBAD/}
catalogue. This is comparable to the value of 1.1 yr$^{-1}$
Gpc$^{-3}$ from \cite{g04}, obtained using a different SFR and a
broken power-law luminosity function.

As GRBs are independent of each other, their distribution is a
Poisson process in time: the probability for at least one event to
occur in the volume out to redshift $z$ during observation time
$T$ at a mean rate $R(z)$ is given by an exponential distribution:
\begin{equation}\label{prob1}
p(n\ge1;R(z),T) = 1 - e^{-R(z) T}\,.
\end{equation}
This formula can be used to define a `probability event
horizon'---as observation time increases, how often will rarer,
more local events,  be observed? See \cite{cow041} for a
description. Based on equation (\ref{prob1}), the probability of
at least one GRB occurring in $z<0.17$ during 4.5 yr is about 0.5,
implying that the observed GRBs in this volume need not be
considered anomalous. But the probability of a GRB occurring in
$z<0.01$ over 4.5 yr is 0.00015, implying that GRB 980425
($z=0.0085$) is either an extreme outlier or a member of a
different GRB population. We model the GRB redshift distribution
under the latter assumption, with the constraint that the
probabilities and rates are consistent with the observed BATSE GRB
distribution.

\section{Simulating an under-luminous GRB redshift distribution}
A GRB distribution comprised of two populations, with different
local rate densities and mean luminosities can be expressed as:
\begin{equation}\label{drdz2}
 \mathrm{d}R =
\frac{\mathrm{d}V}{\mathrm{d}z}\frac{e(z)}{1+z} \Big[\psi_{\rm
c}(z)r^{\rm c}_0+\psi_{\rm u}(z)r^{\rm u}_0\Big] \mathrm{d}z \,,
\end{equation}
where $\psi_{\rm c}(z)r^{\rm c}_0$ and $\psi_{\rm u}(z)r^{\rm
u}_0$ are the scaling functions and local rate densities of the
classical and under-luminous GRBs respectively. It is assumed that
both rates follow the SFR density so that $e(z)$ is the same for
both populations.

Figure 1 plots $\psi_{\rm c}(z)$ and $\psi_{\rm u}(z)$ for BATSE
and {\it Swift} sensitivities. It is evident that at $z>0.1$, the
flux-limit of the detectors severely limits the potential
detectability of the under-luminous GRB population (mean
luminosity $L_{0}=2\times 10^{53}$ s$^{-1}$). For $z<0.1$ the
scaling function is non-negligible even though events will be rare
inside such a relatively small volume. The sensitivity of {\it
Swift} implies that it could potentially detect most
under-luminous events occurring inside a volume bounded by
$z=0.01$.

\begin{figure}
\includegraphics[scale=0.7]{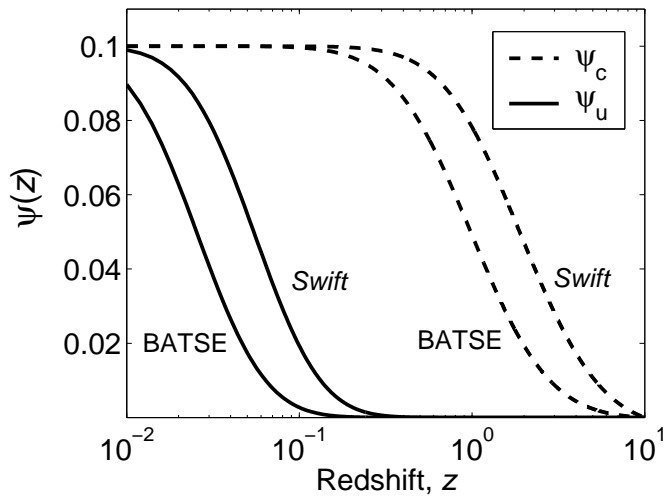} \caption{GRB scaling functions $\psi_
{\rm c}$ for the classical population, and $\psi_{\rm u}$ for an
under-luminous population, shown using BATSE and {\it Swift}
sensitivities. $\psi_{\rm u}$ is calculated using the assumption
that GRB 980425 is a typical event from an under-luminous
population of mean luminosity $L_{0}=2\times 10^{53}$ s$^{-1}$,
three orders of magnitude less than the mean luminosity of the
classical GRBs. A value of 0.1 for the scaling function
corresponds to all GRBs in the field of view of the detector (0.1
str) being potentially detectable.}
\end{figure}

Clearly the observed redshift distribution of GRBs and expected
rates for very energetic Type Ib/c SNe do impose constraints on
the local rate density $r_0^{\rm u}$. Importantly, GRBs 980425 and
031203 show no evidence for jets, implying that there is no
geometric rate enhancement factor required to account for unseen
bursts of similar type. As a first approximation to $r^{\rm u}_0$,
we take the classical rate $r^{\rm c}_0$ increased by the beaming
rate enhancement factor for classical bursts, using a value of 250
from \cite{frail04}. This gives $r^{\rm u}_0\approx 220$ yr$^{-1}$
Gpc$^{-3}$, which is about 0.6\% of the local SN Type Ib/c rate,
$r_0^{\mathrm {SNIbc}}\approx 3.7\times 10^{4}$ yr$^{-1}$
Gpc$^{-3}$ \citep{Izzard04}---a result that supports the view that
SN1998bw was a relatively rare and unusually energetic SN. We note
that the quoted SN Type Ib/c rates may be underestimated because
many core-collapse SN are lost to extinction in most surveys to
date.

The mean of the luminosity function needs to be reduced by 3-4
orders of magnitude and the variance reduced from 2 to 1.5 to
crudely fit the resulting probability distribution with the
observed rates of GRBs at small redshift; that is, the probability
of occurrence of the very nearby GRB 980425 $(z=0.0085)$ must be
compatible with present observations. For this condition to be
satisfied we find that $\sigma$ must be smaller than that used for
modelling the classical population ($\sigma=2$), otherwise the
rate of events at small $z$ would be too large. If a broader
luminosity distribution (larger variance) is employed, for example
$\sigma=2.5$, the mean luminosity must be reduced to 4 orders of
magnitude less than the mean classical GRB luminosity, to yield
observationally consistent probabilities. For definiteness and
consistency, we assume $\sigma=1.5$ and $L_{0}=2\times 10^{53}$
s$^{-1}$ in all calculations

\begin{table}
\caption{The probability of observing at least one GRB inside
volumes bounded by $z=$ 0.0085, 0.05, and 0.17 during 4.5 yr of
observation using a single GRB distribution (classical) with BATSE
sensitivity and a double distribution (classical + under-luminous)
with BATSE and {\it Swift} sensitivities, (labelled as Double
(BATSE) and Double ({\it Swift}) respectively.\label{tbl-1}}
%\begin{ruledtabular}
\begin{tabular}{@{}cccc}
& $p(z<0.0085)$ & $p(z<0.05)$ & $p(z<0.17)$ \\
\hline
Classical (BATSE) & 0.00015 & 0.02 & 0.5 \\
Double (BATSE) & 0.04 & 0.78 & 0.99 \\
Double ({\it Swift})  & 0.04 & 0.96 & 0.99  \\
\hline\\
\end{tabular}
\end{table}

Table 1, using the same parameters as figure 1,  shows the
probabilities for detecting at least one GRB in volumes bounded by
$z=$ 0.0085, 0.05, and 0.17 during 4.5 yrs observation, assuming
the sensitivities of both BATSE/HETE and {\it Swift}. Equation
(\ref{drdz}) is used to calculate the rates based on a classical
distribution and equation (\ref{drdz2}) for a distribution
comprised of both classical and under-luminous bursts.  The double
distribution model increases the detection probability of the very
nearby GRB 980425 ($z=0.0085$) to a still small (0.04), but
significant, level compared with the extremely small probability
(0.00015) from the classical distribution alone.

\begin{figure}
\includegraphics[scale=0.7]{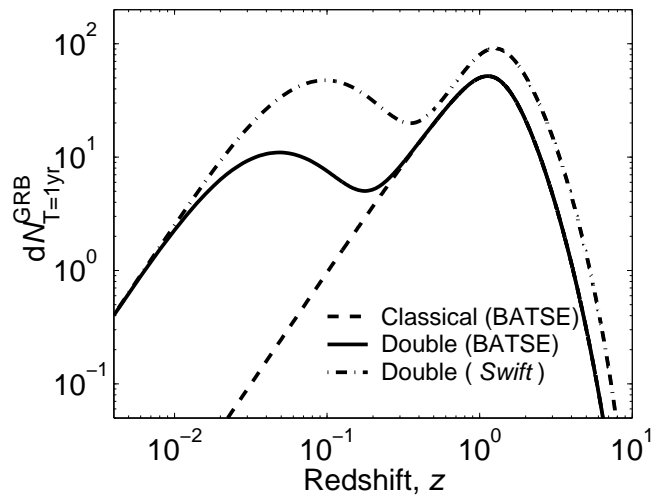} \caption{The differential number of
GRBs as a function of redshift for an observation time of 1 yr
using the three models described in Table 1. The BATSE classical
and double distribution models predict similar numbers from
$z=0.2-10$.}
\end{figure}
\begin{figure}
\includegraphics[scale=0.7]{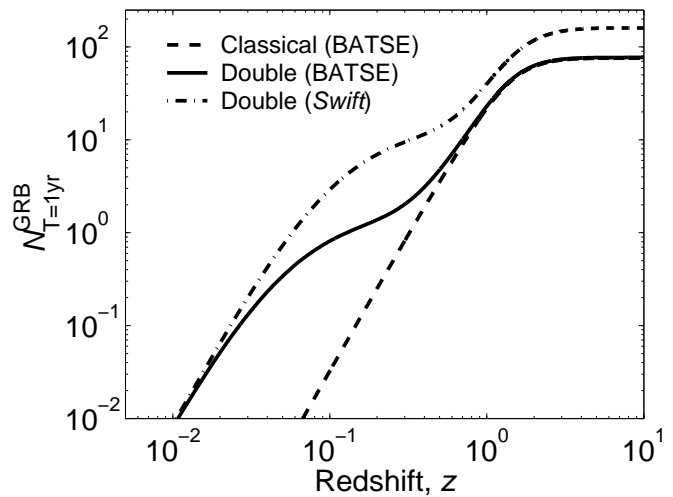} \caption{As for Figure 2 but plotting
the cumulative number of GRBs as a function of redshift for an
observation time of 1 yr. A comparison of the double and classical
models for BATSE shows that the under-luminous population causes a
significant increase in numbers at $z<0.1$, resulting in a higher
probability of observing small-$z$ GRBs. {\it Swift} could
potentially detect over 100 GRBs during one year with up to 5 in
$z<0.1$}
\end{figure}
Figures 2 and 3 plot the number distribution and cumulative number
of GRBs observed over a 1 yr period for the classical and double
distributions assuming BATSE and {\it Swift} sensitivities. For
BATSE sensitivities it is evident that the under-luminous GRBs
have no effect on the observed distribution at $z>0.2$. They do
contribute to the cumulative number at $z<0.2$, a result that is
compatible with the detection of GRBs 980425 and 031203. The
cumulative number increases from about 1 at $z=0.1$, for a BATSE
sensitivity, to about 4 or 5 for a {\it Swift} sensitivity.

\section{Conclusions}
We have shown that a population comprised of classical and
under-luminous GRBs is compatible with the currently observed GRB
redshift distribution that includes the nearby and under-luminous
GRBs 980425 and 031203. We find that a local rate density for an
under-luminous GRB population of $r^{\rm u}_0\approx 220$
yr$^{-1}$ Gpc$^{-3}$, which is about 0.6\% of the local SN Type
Ib/c rate, $r_0^{\mathrm {SNIb/c}}\approx 6.3\times 10^{4}$
yr$^{-1}$ Gpc$^{-3}$, fits the observed low-redshift GRB
distribution. Assuming that GRB 980425 is typical in luminosity
for an under-luminous population, we make a first crude constraint
on such a population by taking a mean luminosity of $L_{0}=2\times
10^{53}$ s$^{-1}$--- about 3--4 orders of magnitude less than the
mean luminosity of the classical GRBs.  These two constraints
yield a probability of BATSE/HETE detecting at least one GRB at
$z<0.05$ to be 0.78 over 4.5 years, a result that is compatible
with the presently observed low-redshift GRB distribution.

GRBs 980425 and 031203 may only appear faint and anomalous because
of the sensitivity limit of BATSE/HETE. If such an under-luminous
population is present, the increased sensitivity of {\it Swift}
should enable it to detect and localize 5 times more GRBs than
BATSE at redshifts $z<0.1$.

It seems reasonable that the observed broad distribution of
observed GRB luminosities may represent related, but different
classes of engine-driven emissions powered by rotating massive
compact stellar remnants. The definite association of some nearby
GRBs with Type Ib/c SNe--a SN type that exhibits considerable
diversity--supports the idea that there could be a diverse class
of inner engines driving at least a fraction of GRBs. These
classes may even form a continuum that encompasses Type Ib/c SNe,
XRF, faint GRBs and classical GRBs. Hence the distribution of
sources in redshift in this scenario would consist of the sum of
the individual populations with different inner engines and local
rate densities. There is most likely overlap between the emission
characteristics of the various sub-populations, so much so that
they may form a continuum of luminosities ranging from XRF to hard
GRBs.

If the anomalous GRBs are all shown to be off-axis observations of
classical GRBs, then the evidence for different classes of cosmic
explosions related to GRBs will become more tenuous. {\it Swift}
should provide new information on the diversity of inner engines
and on the progenitors that produce them, providing a wealth of
data to help solve the cosmic riddle of identifying GRB progenitor
populations.

\section*{Acknowledgments}
I thank the referee for useful suggestions that have led to
improving the clarity of this work. I also thank B. P. Schmidt for
discussing the idea of multiple populations of GRBs and R. R.
Burman and D. G. Blair for providing useful comments. D. M. Coward
is supported by an Australian Research Council fellowship and
grant DP0346344.

\end{document}